# *Shaping the Beam of Light in Nanometer Scales:*

# *A Yagi-Uda Nanoantenna in Optical Domain*

*Jingjing Li, Alessandro Salandrino, and Nader Engheta\**

*University of Pennsylvania*
*Department of Electrical and Systems Engineering,*
*Philadelphia, Pennsylvania 19104*

*Abstract*

A Yagi-Uda-like optical nanoantenna concept using resonant core-shell plasmonic particles as its "reflectors" and "directors" is studied numerically. Such particles when placed near an optical dipole source in a certain arrangement may exhibit large induced dipole moments, resulting in shaping the far-field radiation pattern, analogous to the far field of classical Yagi-Uda antennas in the microwave regime. Variation of the ratio of radii in concentric core-shell nanostructure is used to tailor the phase of the polarizabilities of the particles, and consequently the antenna's far-field pattern.. The idea of a nanospectrum analyzer is also briefly proposed for molecular spectroscopy.

Plasmonic materials have generated increasing interest in various research communities in recent years due to their abilities to manipulate optical signals at deep sub-wavelength dimensions (for a review, see Ref. 1). One of their interesting features is related to the scattering properties of sub-wavelength plasmonic nanoparticles. It is well known that for a particle made of plasmonic materials, the scattering of electromagnetic waves is maximized at certain resonant wavelength determined by the material parameter and the particle geometry, even though the size of the particle may be much smaller than the free-space wavelength. The peculiar characteristics of interaction of light with plasmonic nanoparticles have been known for a long time, and recently, owing to the advancement in nanofabrication technologies, the interest in scattering resonances associated with the presence of plasmonic nanoparticles has been resurrected and explored in detail, both experimentally and theoretically [2-5]. The influence of the plasmonic nanoparticles on the single-molecule spontaneous emission has also been studied, where the stimulated molecule was modeled as a dipole source (see Ref. 6-7 and references there in). These studies have been mostly concentrated on the emission rate of the molecule (i.e., dipole) in a certain environment and only a subset of them has concerned the modification to the radiation pattern (e.g., Refs. 8 and 9).

Design and fabrication of optical nanoantennas has gained growing interest in the recent years [10-11]. It is a challenging task, mainly due to the optical properties of metals that are significantly different from their microwave characteristics. Instead of exploiting their high-conductivity property as in the radio frequency (RF) and microwave domains, their plasmonic features in the optical regime become one of the relevant factors in nano-scale optical antenna design. Motivated and inspired by the various concepts of antenna theory in RF and microwave regimes, such as the conventional RF Yagi-Uda antennas, in a recent work we suggested that plasmonic resonance of core-shell plasmonic particles can be exploited for optical antenna array design [12]. In this idea, a group of resonant plasmonic core-shell nanoparticles is placed around an optical dipole source $p_0$ in a specified pattern. Large induced dipoles may be stimulated on these particles such that the overall far-field radiation characteristics may become different from that of a single dipole source. The optical dipole source can be a stimulated molecule, a quantum dot, or the tip of a near-field scanning optical microscope (NSOM). In the present work, we expand this idea further by studying a Yagi-Uda-like optical nanoantenna and explore this concept using analytical technique and numerical simulations.

The scattering resonance of a concentric plasmonic core-shell spherical nanoparticle has been extensively studied experimentally (e.g., Ref [3]) and theoretically (e.g., Ref [5]). When the size of this particle is much smaller than the incident wavelength, its interaction with the electromagnetic wave is modeled by an induced electric dipole moment with the polarizability $\alpha$ given as $\alpha = -6\pi i \varepsilon_0 c_1 / k_0^3$ where $k_0$ is the free space wavenumber (or the wavenumber in surrounding medium) and $c_1$ the TM Mie scattering coefficient of order 1 (see Ref.8 for more details) whose value is given as $c_1 = -U_1 / (U_1 + iV_1)$ where $U_1$ is

$$U_1 = \begin{vmatrix} j_1(k_1 b) & j_1(k_2 b) & y_1(k_2 b) & 0 \\ \tilde{j}_1(k_1 b)/\varepsilon_1 & \tilde{j}_1(k_2 b)/\varepsilon_2 & \tilde{y}_1(k_2 b)/\varepsilon_2 & 0 \\ 0 & j_1(k_2 a) & y_1(k_2 a) & j_1(k_0 a) \\ 0 & \tilde{j}_1(k_2 a)/\varepsilon_2 & \tilde{y}_1(k_2 a)/\varepsilon_2 & \tilde{j}_1(k_0 a)/\varepsilon_0 \end{vmatrix} \quad (1)$$

and the $e^{-i\omega t}$ convention is assumed. Here, $j_1(x)$ and $y_1(x)$ are the first-order spherical Bessel functions of the first and second kind. $\tilde{j}_1(x)$ stands for $\partial(x j_1(x))/\partial x$ and $\tilde{y}_1(x)$ is similarly defined. $\varepsilon_1$, $\varepsilon_2$ are the permittivity of the core and of the shell, respectively, while $k_1$, $k_2$ are the wavenumbers in each respective region. The outer and inner radii of the particle are denoted as $a$ and $b$. $V_1$ is similar to $U_1$ by replacing the function $\tilde{j}_1$ and $j_1$ in the last column by $\tilde{y}_1$ and $y_1$, respectively. The maximum of the magnitude of $\alpha$ is reached when $V_1$ attains a zero value. A necessary condition for $V_1 = 0$ is to have $\varepsilon_1$ and $\varepsilon_2$ of real values with opposite signs. At a given wavelength for which the imaginary part of the permittivity of the plasmonic material is small compared to the real part, the condition $V_1 = 0$ can be achieved approximately by adjusting $b/a$ (see Fig. 1).

It is well known that for a sub-wavelength uniform sphere made of a given plasmonic material, local plasmonic resonance is achieved when the real part of the relative permittivity (with respect to that of the surrounding medium) is close to -2. However, by adjusting the ratio $b/a$ the scattering resonance can be tailored at other wavelength range for concentric nanoparticle. When such particles, which are placed around an optical dipole source $p_0$, are designed to resonate at the dipole's operating wavelength, the local field inducing the dipole moment on each particle is the superposition of the fields emitting from the original dipole source $p_0$ and all other induced dipoles. Thus the induced dipole $p_i$ on the $i$-th particle can be written as

$$p_i = \alpha_i \left[ \left( \sum_{j \neq i} \Gamma_{ij}(\mathbf{r}_{ij}) \cdot \mathbf{p}_j \right) + \Gamma_{i0}(\mathbf{r}_{i0}) \cdot p_0 \right] \qquad (2)$$

where $\mathbf{r}_{ij}$ is the position vector of particle $i$ with respect to particle $j$, $\Gamma_{ij}(\mathbf{r}_{ij})$ is the dyadic Green function of the dipole $p_j$ evaluated at the position of the $i$-th particle, which is known when the positions of all the particles and the source dipole are given. The original optical dipole source is expressed by $p_0$, and its field on dipole $i$ is written separately as the second term in the right hand side. This system of linear equations can be solved directly. The far-field radiation pattern of the system thus is equivalent to that of the $N + 1$ dipole system — the source dipole and $N$ induced dipoles — with the magnitude and phase of each induced dipole determined from Eq. (4). An example is shown in Fig. 2. For this system two identical concentric plasmonic particles are placed at each side of the optical dipole $p_0$ at a distance of $0.35\lambda_0$, with a free space wavelength of $\lambda_0 = 620 nm$, as in Fig. 2(a). The cores of the particles are assumed to be made of SiO2 with $\varepsilon_1 = 2.2\varepsilon_0$, the shells are made of silver with $\varepsilon_2 = (-15.33 + 0.451i)\varepsilon_0$ at this operating wavelength, and $a = 0.1\lambda_0$, $b/a = 0.842$. For these parameters the polarizability is $\alpha = 0.063 e^{i0.48\pi} \varepsilon_0 \lambda_0^3$. The pattern of radiated power is given in (b) and (c) which are normalized to the maximum point of radiated power flux density of the original source dipole when it radiates in the absence of particles. It can be clearly seen that, as expected both the magnitude and the shape of the pattern are modified by the presence of the particles when compared with the pattern of the original dipole source alone. In this sense, the plasmonic particles take the role of an antenna array at optical wavelengths.

In the field of antenna design, it is commonly required to design an antenna system with a prescribed far-field radiation pattern. For optical nanoantennas using plasmonic particles as described above, the induced dipole moment on each particle is resulted from the fields of the source and all the other induced dipoles. This implies that all the induced dipoles are essentially coupled to one another such that we do not have an arbitrarily separate control over the phase and magnitude of each dipole element individually. This makes the pattern synthesis problem for the collection of plasmonic particles in the optical antenna array a challenging, but still achievable task. An interesting analogous case of such a pattern synthesis in the microwave domain is the celebrated Yagi-Uda antenna [13]. In this design, as shown in Fig.4(a), a half-wavelength resonant dipole antenna is driven by a source (e.g., a voltage source), and several other wire elements with proper lengths performing as passive elements, are placed at specific positions in proximity of the driven dipole, in order to re-shape the far-field radiation pattern. One of these passive wire elements longer than the resonating length is placed at one side of the source antenna. This element with an inductive impedance response works as a "reflector". Several other wire elements with lengths shorter than the resonating length are situated at the other side of the driven source. These shorter elements, which are called directors, have capacitive impedance response. The resulting radiation pattern of such an antenna has a narrow beam towards the direction of the "directors" and a minimum (or a null) towards the "reflector" direction. The essential idea in the classic Yagi-Uda antenna was to introduce some break in the symmetry of the response phases of the scatterers placed at each side of the source so that the shape of the total radiation distribution exhibits certain break in the symmetry and more directivity. Such an idea may be transplanted into our optical nanoantenna array design where the core-shell nanoparticles can play the roles of "reflectors" and "directors". Ordinarily, the values of ratio of radii *b/a* for the core-shell nanoparticle are designed to set the particles at scattering resonance. Under this condition the phase of the polarizability is close to $\pi/2$ and the equivalent current of the induced dipole is almost in phase with the incident field. However, *b/a* can be designed deliberately to "de-tune" this resonance in order to achieve a phase of induced dipole less than or greater than $\pi/2$ compared to the incident field. Therefore, these detuned nanoparticles may play the role of the "reflector" or the "director" element in the Yagi-Uda antenna. By placing one "reflector" nanoparticle at one side of the original dipole source, and one or several "director" nanoparticles at the opposite side, we can design a Yagi-Uda antenna at optical wavelengths.

To test our idea, we first show a design of two-particle Yagi-Uda antenna working at 620*nm*. Both particles have the core with SiO2 and the shell with silver as those in Fig. 2. The outer radius of each particle is still $0.1\lambda_0$. However, this time for the particle at the left we use *b/a* =0.851 so that $\alpha_1 = 0.0592 e^{i0.6\pi} \varepsilon_0 \lambda_0^3$, while the one at the right we use *b/a* =0.834 so that $\alpha_2 = 0.0611 e^{i0.4\pi} \varepsilon_0 \lambda_0^3$. Notice the phases of the polarizabilities are shifted to be above or below $\pi/2$. The two particles are placed at equal distance *d* from the source dipole. Then *d* is varied in order to achieve a "proper" radiation pattern, which should have a small value at $0^o$ or $180^o$ and a large value at the opposite. The pattern of one of these designs is shown as the bold solid line (red online) in Fig. 3, for which *d* =$0.14\lambda_0$. For this case, the radiation is maximized along $180^o$ and minimized along the opposite direction. We declare the particle at the right side a capacitive element because the effective current of the dipole ($I_0 l = -i\omega p$) has a leading phase compared to the incident field. For a similar reason, the particle at the left is called an inductive one. In this design, the inductive particle (the left one) works as the director while the capacitive one (the right side one) as the reflector. Since the detailed situation is different from that of the classic Yagi-Uda antenna at RF and microwave frequencies, the reflector may not necessarily be the capacitive element, nor is the director necessarily the inductive element. Under other values of *d*, for example, $d = 0.25\lambda_0$, there could be a radiation maximum towards the right while relatively small radiation towards the left, as shown by the bold dashed line (blue online) in Fig. 3.

However, the conceptual idea is analogous, that is, by introducing some asymmetry in the two elements placed at the opposite sides of the source dipole, it may be possible to shape the beam patterns accordingly. The pattern varies with wavelength rapidly, since the polarizabilities of each particle are sensitive to the operating wavelength when their geometries (the outer and inner radii) are already chosen. Also, at different wavelengths the relative distances between the source dipole and the particles are electrically different (while they are physically fixed), and thus the coupling among the particles is changed with wavelength. As the operating wavelength shifts away from the designed value (here 620nm) the radiation magnitude decreases, and the pattern no longer has the dominant radiation beam along $180^o$ (not shown here). This wavelength sensitivity of the beam patterns can offer exciting potential applications, one of which we will briefly mention later here.

Yagi-Uda optical antennas with more than one director can also be designed in order to achieve narrower beam pattern. In this case, several degrees of freedom are involved and we need to optimize the antenna design heuristically. Suppose we want to design a one-reflector, eight-directors Yagi-Uda optical antenna. One of the design processes begins from the 2-particle Yagi-Uda optical antenna shown in Fig.3 with the particle-to-dipole-source distance as $d = 0.25\lambda_0$, for which the $0.6\pi$-phase particle works as a reflector and the $0.4\pi$-phase particle works as a director. To design a multiple-director Yagi-Uda optical antenna, we use the same reflector, and fix the reflector-source distance to be $d_1 = 0.25\lambda_0$. Suppose the distances between each directors are the same, and is equal to the distance between the dipole source and the first director $d_2$. $d_2$ is then varied to achieve a "proper" pattern. The pattern of a "good" design for $d_2 = 0.65\lambda_0$ is shown in Fig.4 as the bold dashed line (blue online). Clearly, a narrower beam is achieved compared to that in Fig. 3. Later when studying the frequency property of this optical antenna, we noticed that actually such a system yields an even narrower beam at a slightly different working wavelength $\lambda_0 = 646nm$, as shown in Fig. 4 by the bold solid line (red online). At this frequency, the phase of the polarizability of the "reflector" particle is $0.45\pi$ and those of the eight "directors" are $0.30\pi$. Such a pattern is much more attractive than that of Fig. 2 or Fig. 3 when the directivity is concerned.

In the molecular florescence, the emission spectrum may often cover a wide wavelength range, thus a stimulated molecule may emit as a broad-band source. We can design a set of more than one Yagi-Uda antenna array driven by the same molecular source, but each optimized at different wavelengths such that the maximum beam at a given wavelength points to a different direction. This can essentially lead to analyzing the spectral information of the molecular emission, transforming the spectral contents into angular variation of emitted signals, and thus effectively providing a "nano spectrum analyzer" at the nanometer scale at optical wavelengths. The details of our analysis for such a device, which can have various potential applications in molecular spectroscopy and sensing, will be given in a future publication.

This work is supported in part by the U.S. Air Force Office of Scientific Research (AFOSR) grant number FA9550-05-1-0442.

* To whom correspondence should be addressed: E-mail: engheta@ee.upenn.edu

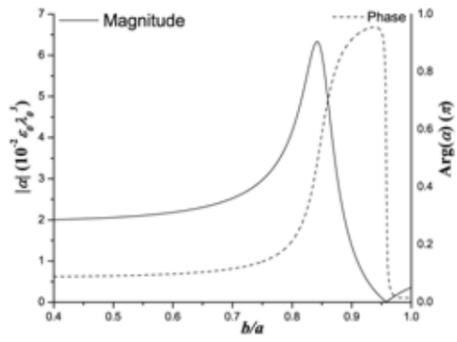

Fig.1 Magnitude and phase of polarizability $\alpha$ vs $b/a$ of a concentric particle. Operating wavelength is 620nm. The core is made of SiO$_2$, ($\varepsilon_1 = 2.2\varepsilon_0$) and shell of Silver ($\varepsilon_2 = (-15.33 + 0.451i)\varepsilon_0$).

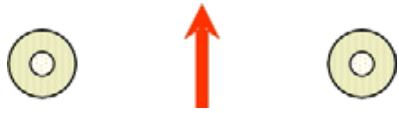
(a)

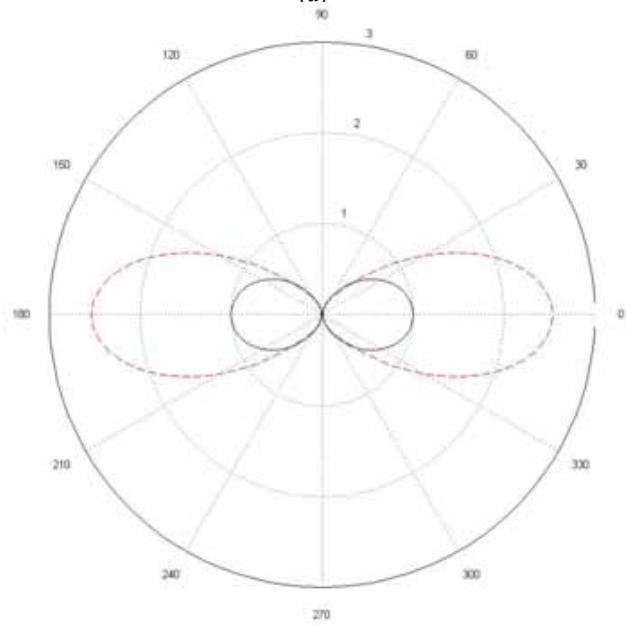
(b)

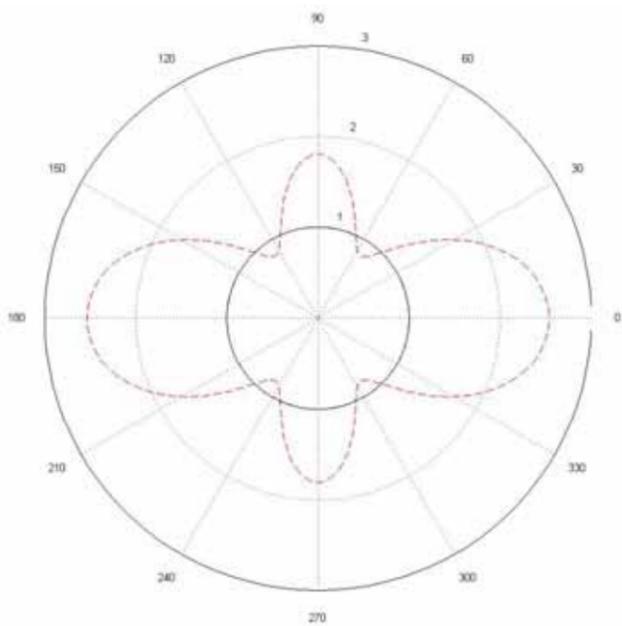
(c)

-12

FIG. 2(Color online): An optical antenna array composed of two identical plasmonic core-shell nanoparticles placed besides an optical dipole source. (a): Geometry of the problem. (b): Pattern of the radiated power flux density in the E-plane and (c) in the H-plane. In the patterns the dotted (red) curve is that of the antenna array, and the solid (black) curve is that of a single dipole alone. The patterns are normalized with respect to the maximum of the radiated power flux density of the original dipole source when it radiates in the absence of the particles.

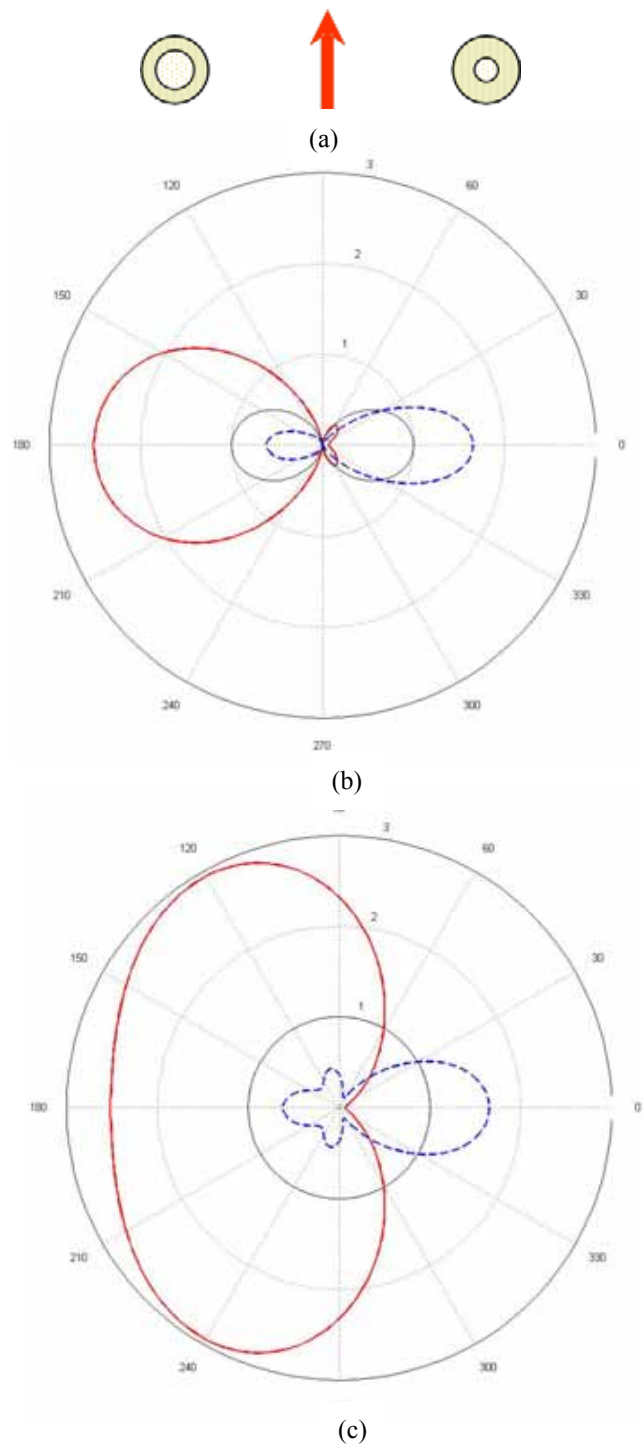

Fig. 3(Color online): A 2-particle optical Yagi-Uda nanoantenna operating at 620nm. (a): The geometry. (b): Pattern of the radiated power flux density in the E-plane and (c) in the H-plane. The different lines in (b)

---

-14

and (c) are the patterns for different dipole-to-particle distances. Bold solid (red online): $d = 0.14\lambda_0$; Bold dashed (blue online): $d = 0.25\lambda_0$; light solid (black online): a single dipole alone. The patterns are normalized in the way similar to those in Fig. 2

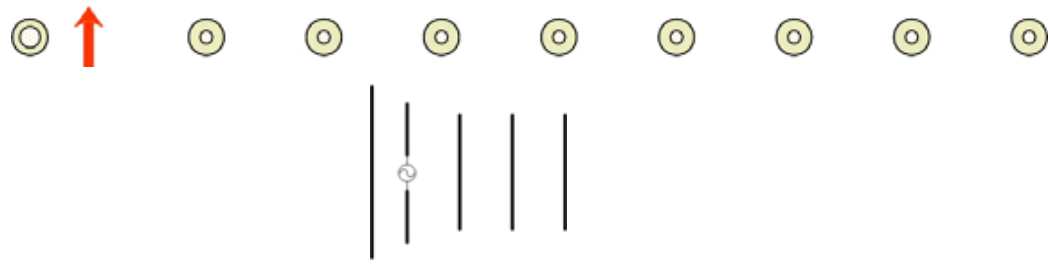

(a)

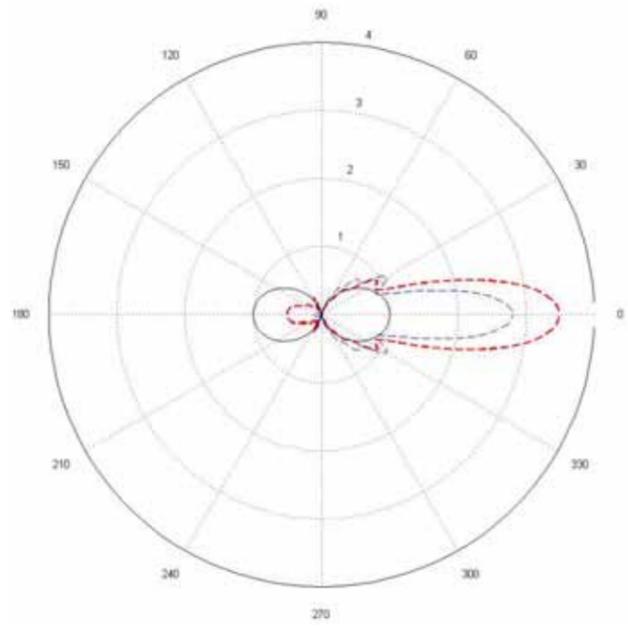

(b)

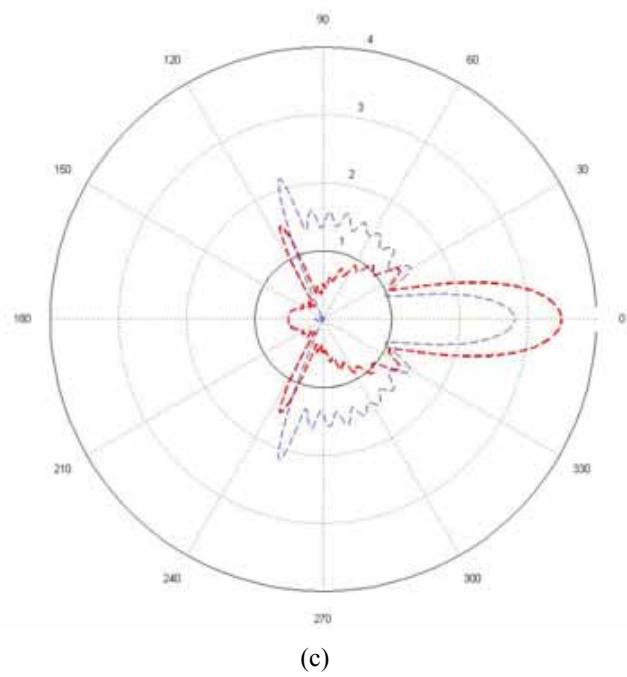

(c)

Fig. 4(Color online): A 9-particle optical Yagi-Uda nanoantenna (a): The geometry of the optical Yagi-Uda antenna and the diagram of a classical Yagi-Uda antenna in the RF and microwave domains. (b): Pattern of the radiated power flux density in the E-plane and (c) in the H-plane. The light dash line (blue online): the pattern at 620nm; The bold dash line (red online): the pattern at 645nm. The solid (black online): the pattern of the original dipole source alone.